\documentstyle[twoside,fleqn,espcrc2,epsf]{article}

\newcommand{\AmS}{{\protect\the\textfont2
  A\kern-.1667em\lower.5ex\hbox{M}\kern-.125emS}}

\title{
Aspects of the thermal phase transition of QCD with 
small chemical potential
\thanks{Presented by S. Ejiri. 
Numerical work was performed using the UKQCD APEmille in Swansea 
supported by PPARC grant PPA/a/s/1999/00026, 
and in part by EU contract ERBFMRX-CT97-0122.
}}

\author{ S. Ejiri\rlap,\address{Department of Physics, 
University of Wales Swansea, Singleton Park, Swansea, SA2 8PP, U.K.} 
C.R. Allton\rlap,$^{\rm a}$ S.J. Hands\rlap,$^{\rm a}$ 
O. Kaczmarek\rlap,\address{Fakult\"{a}t f\"{u}r Physik, 
Universit\"{a}t Bielefeld, D-33615 Bielefeld, Germany} 
F. Karsch\rlap,$^{\rm b}$ E. Laermann\rlap,$^{\rm b}$
and L. Scorzato$^{\rm a}$}

\begin{document}

\begin{abstract}
We propose a new method to investigate the thermal properties of QCD 
with a small chemical potential $(\mu)$. 
The derivatives of the phase transition point with respect to $\mu$ 
are computed for 2 flavors of p4-improved staggered fermions. 
We moreover comment on the complex phase of the fermion determinant 
in finite-density QCD.
\end{abstract}

\maketitle

\section{Introduction}
\label{sec:intro}

The study of the quark-gluon plasma in heavy-ion collision experiments 
is one of the most interesting topics in contemporary physics.
To understand these experiments, precise theoretical inputs of 
the phase transition of QCD are indispensable.
Over the last several years, the numerical study of lattice QCD 
has been successful at zero quark chemical potential $(\mu)$ and 
high temperature \cite{Eji00}.
In contrast, because the quark determinant is complex at $\mu \neq 0$ 
and Monte-Carlo simulation is not applicable directly, study 
at finite $\mu$ is still in the stage of development.

Fortunately, the interesting regime for heavy-ion collisions is 
rather low density, e.g., $\mu \sim 15 {\rm MeV}$ $(\mu /T_c \sim0.1)$ 
for RHIC \cite{Redl01}. 
Therefore, an approach of Taylor expansion may be the most efficient way, 
by computing the derivatives of physical quantities in terms of $\mu$ 
at $\mu=0$ to determine these coefficients. 
Pioneering works in such a framework are given for free energy 
(quark number susceptibility)\cite{Gott88} and screening mass\cite{Taro01}.

In this study, we investigate the transition point $(\beta_c)$ at 
finite $\mu.$ A similar study has been done by \cite{Fod01}.
In sec.\ \ref{sec:method}, we propose a new method to compute 
the derivative of the physical quantities with respect to $\mu.$ 
The result of the second derivative of $\beta_c$ for 2 flavor QCD 
is given in sec.\ \ref{sec:simulation}. 
We discuss the problem of the complex phase 
in sec.\ \ref{sec:phase}. Section \ref{sec:summary} presents the conclusions.

\section{Reweighting method for $\mu$-direction}
\label{sec:method}

Basically most of the attempts to study finite density QCD have been 
tried by performing simulations at $\mu =0$ and reweighting by the identity: 
\begin{eqnarray}
\label{eq:rew}
& \hspace{-3mm} \langle {\cal O} \rangle_{(\beta, \mu)} & \hspace{-3mm} 
= \left\langle {\cal O} W \right\rangle_{(\beta_0,0)} /
\left\langle W \right\rangle_{(\beta_0,0)}, \hspace{10mm} \\
 &\hspace{-3mm} W =& \hspace{-4mm} {\rm e}^{\alpha N_{\rm f}
(\ln \det M(\mu) - \ln \det M(0))} {\rm e}^{-S_g(\beta)+S_g(\beta_0)}, 
\nonumber
\end{eqnarray}
where $M$ is the fermion matrix, $S_g$ is the gauge action, 
$N_{\rm f}$ is the number of flavors and 
$\alpha$ is 1 or $1/4$ for the Wilson or staggered fermion.

The reweighting factor of the gauge part is easy to compute \cite{Swen88}.
However, the fermion part is highly non-local and difficult to 
compute in practice. 
Here we consider a method which is applicable for small $\mu$.
We perform a Taylor expansion for the fermion part of 
the reweighting factor around $\mu=0$.
Similarly, we expand fermionic observables, e.g., the chiral condensate, 
$\langle \bar{\psi} \psi \rangle = 
(1/V) \alpha N_{\rm f} \langle {\rm tr} M^{-1} \rangle$.
If we consider up to the $n$-th derivative of both the reweighting factor 
and the fermionic observable, we can calculate the $n$-th derivative of 
the physical quantity with respect to $\mu$ correctly, 
which can be easily checked by performing a Taylor expansion of 
each expectation value of the physical quantity.
The derivatives of $\ln \det M$ and ${\rm tr} M^{-1}$ can be calculated 
by the random noise method, which enables us to compute on a rather large 
lattice in comparison with usual studies of finite density QCD.

Moreover, we should notice that the odd order derivatives of 
$\ln \det M$ and ${\rm tr} M^{-1}$ are pure imaginary and 
the even order derivatives are real at $\mu = 0.$
This property is proved using the identities:
$M^{\dagger}(\mu) = \gamma_5 M(-\mu) \gamma_5,$ and
$\frac{{\rm d}^n M^{\dagger}}{{\rm d} \mu^n}(\mu) = 
(-1)^n \gamma_5 \frac{{\rm d} M}{{\rm d} \mu}(-\mu) \gamma_5.$
From this property, we can explicitly confirm that, 
if the operator we compute is real at $\mu=0$, e.g., 
$\langle \bar{\psi} \psi \rangle$, the first derivative of 
the expectation value is zero at $\mu=0$, 
as we expect from the symmetry under the change from $\mu$ to $-\mu$.
We moreover can easily estimate the phase factor of the fermion determinant 
from this feature, which is discussed in sec.\ \ref{sec:phase}.

\section{Results for $N_{\rm f}=2$ improved staggered}
\label{sec:simulation}

\begin{table}[tb]
\caption{}
\label{tab:betac}
\vspace*{-6mm}
\begin{center}
\begin{tabular}{ccccc}
\hline
$m$ & $\beta_c(\bar{\psi}\psi)$ & ${\rm d}^2 \beta_c / {\rm d} \mu^2$ &
 $\beta_c$(Polyakov) & ${\rm d}^2 \beta_c / {\rm d} \mu^2$ \\
\hline
0.1 & 3.6463(25) & $-$1.41(61) & 3.6489(27) & $-$1.71(67) \\
0.2 & 3.7575(19) & $-$1.51(42) & 3.7590(20) & $-$1.61(36) \\
\hline
\end{tabular}
\vspace*{-4mm}
\end{center}
\end{table}

\begin{figure}[t]
\vspace*{-7mm}
\centerline{
\epsfxsize=6.1cm\epsfbox{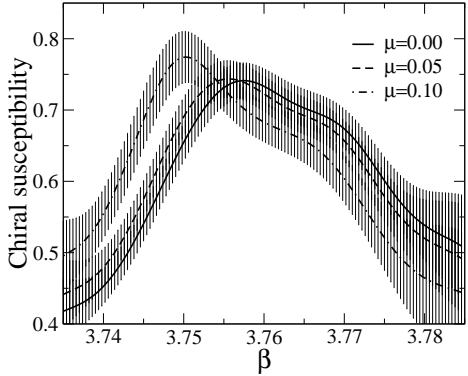}
}
\vspace*{-11mm}
\caption{
Chiral susceptibility at $m=0.2$.
}
\vspace*{-4mm}
\label{fig:csus02}
\end{figure}

\begin{figure}[t]
\vspace*{-0mm}
\centerline{
\epsfxsize=6.2cm\epsfbox{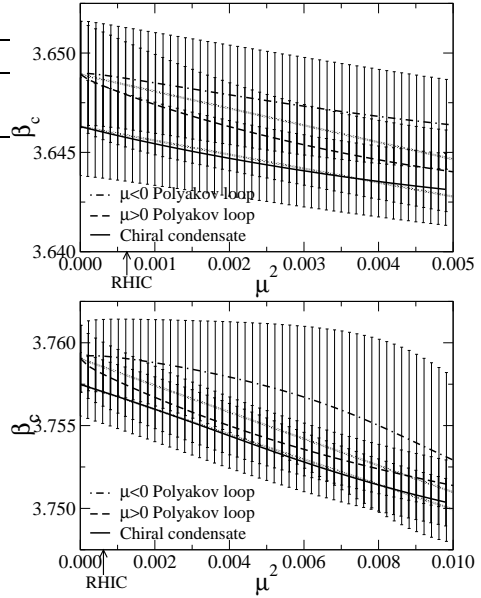}
}
\vspace*{-10mm}
\caption{
Phase transition points for $m=0.1$ (upper) and $0.2$ (lower).
Gray lines are the fitted lines.
}
\vspace*{-4mm}
\label{fig:betac}
\end{figure}

We calculate the second derivative of $\beta_c$ with respect to $\mu$. 
We employ a combination of the Symanzik improved gauge and 
the p4-improved staggered fermion actions \cite{p4action}. 
It is known that this action makes the discretization error of pressure 
small at high temperature, and $T_c$ obtained by this action is 
consistent with that using improved Wilson fermions \cite{CPPACS00}.
To include the finite density effect, we multiply the hopping terms 
proceeding $n$-steps in positive and negative temporal directions 
by ${\rm e}^{n \mu}$ and ${\rm e}^{-n \mu}$ respectively.

We investigate the transition points at $m=0.1$ and $0.2$. 
The corresponding pseudo-scalar and vector meson mass ratios are 
$m_{PS}/m_{V} \approx 0.7$ and $0.85$. 
We compute Polyakov loop, chiral condensate, and their susceptibilities. 
The simulations are performed on a $16^3 \times 4$ lattice 
at $\beta=3.64$-$3.65$ with a total of 45000 trajectories for $m=0.1$, 
and $\beta=3.74$-$3.80$ with 76700 (Polyakov) or 69000 $(\bar{\psi}\psi)$ 
trajectories for $m=0.2$. 
10 sets of Z(2) noise vectors are used for each trajectory 
to compute the reweighting factor up to the second order.
We plot, in Fig.\ref{fig:csus02}, the chiral susceptibility at $m=0.2$. 
This figure shows that the peak position becomes lower as $\mu$ increases. 
Figures \ref{fig:betac} show the transition point determined 
by the peak position of the Polyakov loop susceptibility and 
the chiral susceptibility as a function of $\mu^2$. 
Because the first derivative is zero, as we discussed before, 
we fit the data for $\beta_c$ by a straight line in $\mu^2$, 
fixing $\beta_c$ at $\mu=0$, in a range, $\mu^2 \leq 0.005$ 
and $0.01$ for $m=0.1$ and $0.2$, respectively, 
in which the phase problem is not serious (see sec.\ \ref{sec:phase}). 
We then obtain the results in Table \ref{tab:betac}. 
Quark mass dependence of ${\rm d}^2 \beta_c/{\rm d} \mu^2$ is 
not visible within the error.
We denote the interesting value of $\mu$ for RHIC in Figs.\ref{fig:betac}. 
The shift of $\beta_c$ from $\mu=0$ is found to be small at this point.

The second derivative of $T_c$ can be estimated by 
$\frac{{\rm d}^2 T_c}{{\rm d} \mu_{\rm phys}^2} = -\frac{1}{N_t^2 T_c} 
\left. \frac{{\rm d}^2 \beta_c}{{\rm d} \mu^2} \right/ 
\left( a \frac{{\rm d} \beta}{{\rm d} a} \right).$
We obtain the beta-function from the data of the string tension 
in Ref.\cite{p4action},
$a^{-1} ({\rm d} a / {\rm d} \beta) = -2.08(43)$ at 
$(\beta, m) =(3.65, 0.1)$. We then find  
$ T_c ({\rm d}^2 T_c / {\rm d} \mu_{\rm phys}^2) \approx -0.2$ at $m=0.1$.
The curvature of the transition line in the small $\mu$ 
region seems to be a little smaller than the phenomenological expectation, 
a tendency also suggested by the result of Fodor and Katz \cite{Fod01}.

\section{Comment on the phase (sign) problem}
\label{sec:phase}

\begin{figure}[t]
\vspace*{-0mm}
\centerline{
\epsfxsize=6.1cm\epsfbox{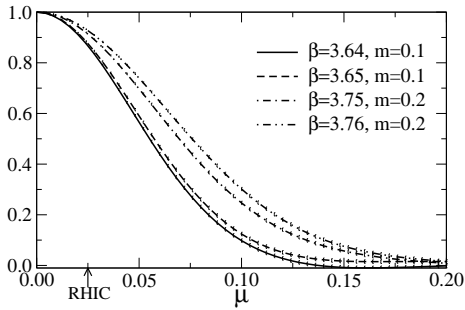}
}
\vspace*{-10mm}
\caption{
Expectation value of $\cos\theta$ near $\beta_c$ 
as a function of $\mu$ on a $16^3 \times 4$ lattice.
}
\vspace*{-4mm}
\label{fig:phase}
\end{figure}

If the reweighting factor in Eq.(\ref{eq:rew}) changes its sign frequently 
due to the complex phase of the quark determinant, 
the calculation of the expectation value becomes very difficult, 
which hinders the study at $\mu \neq 0$. 
The phase of the quark determinant is zero at $\mu=0$ and 
becomes larger as $\mu$ increases. Therefore, 
it is important to consider for which value of $\mu$ the problem 
becomes severe.

As discussed in section \ref{sec:method}, the phase can be expressed 
using the odd terms of the Taylor expansion of $\ln \det M$. 
Denoting $\det M = |\det M| {\rm e}^{i \theta},$
$
\theta = \alpha N_{\rm f} \ {\rm Im} \left[\mu \frac{{\rm d} 
\ln \det M}{{\rm d} \mu} 
+ \frac{\mu^3}{3!} \frac{{\rm d}^3 \ln \det M}{{\rm d} \mu^3} 
+ \cdots \right]. \nonumber
$
For small $\mu$, the first term, $\alpha N_{\rm f} {\rm Im} 
\frac{{\rm d} \ln \det M}{{\rm d} \mu} \mu = 
\alpha N_{\rm f} {\rm Im} \ {\rm tr} \left( M^{-1} 
\frac{{\rm d} M}{{\rm d} \mu} \right) \mu,$ is dominant. 
Because the average of the phase factor is zero, the phase fluctuation 
is important. We investigate the standard deviation of 
$\frac{1}{V} {\rm Im} \ {\rm tr} \left( M^{-1} \frac{{\rm d} M}{{\rm d} \mu} 
\right)$. 
We obtain that the magnitude of the phase fluctuation is 
about $21 \mu$ for $m=0.1$ and $16 \mu$ for $m=0.2$ near $\beta_c$. 
Consequently the phase problem starts to appear at $\mu \sim 0.07 (0.1)$, 
i.e., $\mu_{\rm phys}/T_c \sim 0.3 (0.4)$ for $m=0.1 (0.2)$, 
since the phase problem arises if the phase fluctuation becomes larger 
than $O(1)$. We notice that the value of $\mu$ for which the phase 
fluctuations becomes significant decreases as the quark mass decreases. 
Figure \ref{fig:phase} shows the average of the phase factor 
${\rm e}^{i \theta}$. We can see that the average becomes small 
around those values of $\mu$.

\section{Conclusions}
\label{sec:summary}

We propose a new method based on Taylor expansion to investigate thermal 
properties of finite density QCD. 
We computed the chiral susceptibility and the Polyakov loop susceptibility 
for 2 flavors of p4-improved staggered fermions, and found that 
$\beta_c$ and $T_c$ become smaller as $\mu$ increases. 
Our results suggest that the discrepancy of $T_c$ from $\mu=0$ is 
small in the interesting region for heavy-ion collisions. 
We also estimated the complex phase of the fermion determinant 
for a $16^3 \times 4$ lattice. We found that the sign problem is not 
serious in the range of $\mu_{\rm phys}/T_c < 0.3$-$0.4$ for $m=0.1$-$0.2$, 
which covers the regime of the heavy-ion collisions at RHIC.

\end{document}